# $^{29}$Si Hyperfine Structure of the E'$_{\alpha}$ Center in Amorphous Silicon Dioxide

G. Buscarino*, S. Agnello and F. M. Gelardi

Department of Physical and Astronomical Sciences, University of Palermo, Via Archirafi 36, I-90123 Palermo, Italy

We report a study by electron paramagnetic resonance (EPR) on the E'$_{\alpha}$ point defect in amorphous silicon dioxide (a-$SiO_2$). Our experiments were performed on γ-ray irradiated oxygen-deficient materials and pointed out that the $^{29}$Si hyperfine structure of the E'$_{\alpha}$ consists in a pair of lines split by ~49 mT. On the basis of the experimental results a microscopic model is proposed for the E'$_{\alpha}$ center, consisting in a hole trapped in an oxygen vacancy with the unpaired electron sp$^3$ orbital pointing away from the vacancy in a back-projected configuration and interacting with an extra oxygen atom of the a-$SiO_2$ matrix.

PACS numbers: 61.72.Ji, 61.18.Fs, 61.43.Dq, 61.80.Ed

The most important and studied radiation-induced point defects in amorphous silicon dioxide (a-$SiO_2$) are the E' centers [1,2]. Four types of these paramagnetic point defects, E'$_{\alpha}$, E'$_{\beta}$, E'$_{\gamma}$ and E'$_{\delta}$, have been distinguished in bulk materials on the basis of their electron paramagnetic resonance (EPR) line shapes, phenomenological production and thermal stability [1-3].

The E'$_{\gamma}$ center is commonly observed in room temperature irradiated a-$SiO_2$. It is characterized by an almost axially symmetric EPR main line with a zero crossing g value of ~2.0006, ad by a pair of EPR lines split by ~42 mT, arising from the hyperfine interaction of the unpaired electron with a $^{29}$Si nucleus (nuclear spin I=1/2, natural abundance 4.7%) [1-3]. The most accepted model of the E'$_{\gamma}$ center consists in a puckered positively charged oxygen vacancy: O≡Si$^{\bullet}$ O≡Si$^{+}$ (where ≡ represents the bonds to three oxygen atoms, $^{\bullet}$ is an unpaired electron and $^{+}$ is a trapped hole) [1-3]. In this model it is supposed that, following ionization of the vacancy, the positively charged Si atom moves backward through the plane of its basal oxygens in a puckered configuration. As a consequence of this structural relaxation, the unpaired electron localizes in an sp$^3$ hybrid orbital of the unpuckered Si atom [1-3]. However, on the basis of experimental [4] and theoretical [5] investigations, it has been also suggested that after hole capture by an oxygen vacancy the Si atom on which the unpaired electron is localized could relax backward through the plan of its three basal oxygens atoms in a back-projected configuration.

The E'$_{\beta}$ center has very similar spectroscopic features with respect to that of the E'$_{\gamma}$, but it is believed to originate from the interaction of hydrogen atoms with some precursor site in a-$SiO_2$ [1-3].

The E'$_{\delta}$ center is characterized by a nearly symmetric EPR main resonance with g~2.002 and a $^{29}$Si hyperfine doublet split by ~10 mT [1-3,6-9]. Since its first observation [6] the microscopic model of this point defect has been largely debated [9]. However, a recent experimental investigation [7] has pointed out that the unpaired electron wave function of this defect is delocalized over four nearly equivalent Si atoms, supporting a microscopic model of the E'$_{\delta}$ center consisting in an unpaired electron delocalized over a pair of nearby oxygen vacancies or over a 5-Si cluster in a-$SiO_2$ [7-9].

The E'$_{\alpha}$ center, at variance to the other E' centers, posseses an orthorhombic g matrix [1,3]. Griscom first observed this defect in high purity (stoichiometric) a-$SiO_2$ materials X-ray irradiated at 77 K [3]. In that work the author found that the E'$_{\alpha}$ is stable only below T≅200 K and that it converts to the E'$_{\gamma}$ [10] upon exposure to room light [3]. In a successive experimental investigation, Griscom and Friebele [6] observed the E'$_{\alpha}$ defect in nonstoichiometric a-$SiO_2$ γ-ray irradiated at room temperature. However, at variance to the first experimental investigation, the E'$_{\alpha}$ defect was found to be thermally stable at room temperature and no light induced conversion to E'$_{\gamma}$ was observed [6].

In 1984 Griscom suggested [3] that the $^{29}$Si hyperfine structure of the E'$_{\alpha}$ center consists in a doublet split by ~42 mT, as for E'$_{\gamma}$ and E'$_{\beta}$ centers. On the basis of this attribution, he proposed that the E'$_{\alpha}$ center could originate by an irradiation induced displacement of an O atom and an electron from a regular site of the a-$SiO_2$ matrix. This generates a positively charged oxygen vacancy similar to the one involved in the E'$_{\gamma}$ center, but for the displaced O atom forming a peroxy bridge with one of the basal O atoms of the O≡Si$^{\bullet}$ moiety [3]. The perturbation induced by the knocked O atom on the unpaired electron wave function being responsible for the orthorhombic g matrix of the E'$_{\alpha}$ center [3]. In 2000, after a reexamination of the previously published data, Griscom proposed [1] that the $^{29}$Si hyperfine structure of the E'$_{\alpha}$ center should consist in a pair of lines split by ~13 mT. On the basis of this latter attribution, a model consisting in a twofold coordinated Si (O=Si:, where : represents a lone pair) having trapped an electron was put forward [1].

Successively, Uchino *et al.* [11] on the basis of quantum-chemical calculations have suggested that the E'$_{\alpha}$ center could originate from an hole trapped in a twofold coordinated Si. The authors showed that upon hole capture the structure relaxes to a metastable state characterized by a sp$^3$-like unpaired electron wave function and with an expected value of the isotropic hyperfine coupling constant of ~44 mT [11]. Furthermore, the system was found to easily relax to a more stable structure in which an O≡Si• moiety is formed, giving a possible explanation of the experimentally observed room light induced conversion from $E'_\alpha$ to $E'_\gamma$ [3,11].

Until now a general consensus on the actual microscopic structure of the $E'_\alpha$ center has been not yet reached. However, as it comes from the above discussion, a way to establish a definitive structure for the $E'_\alpha$ center could result by the definitive identification of its $^{29}$Si hyperfine structure.

In the present Letter, we report on the generation of $E'_\alpha$ centers in oxygen-deficient materials. Our study has permitted us to point out that the $^{29}$Si hyperfine structure of the $E'_\alpha$ center actually consists in a pair of lines split by ~49 mT. On the basis of the experimental results a microscopic structure is proposed for the $E'_\alpha$ center.

The materials considered here are commercial a-$SiO_2$. Three of these are obtained from fused quartz, KI [12], QC and Pursil 453 [13], while a fourth material, KUVI [12], is synthesized by vapour axial deposition technique. One sample of each material has been γ-ray irradiated at room temperature at a dose of ~$10^2$ kGy. In the irradiated samples we have estimated the concentration of $[AlO_4]^0$ centers by EPR measurements, whereas that of oxygen vacancies and twofold coordinated Si were determined by measuring the amplitude of the 7.6 eV optical absorption band and of the luminescence band peaked at ~4.4 eV excited at ~5 eV, respectively [14]. In all the considered materials we have estimated a comparable concentration of $[AlO_4]^0$ centers of ~$10^{17}$ spins/cm$^{-3}$, but they were found to differ in the oxygen deficiency. In KUVI, QC and Pursil 453 a concentration of oxygen vacancies higher than ~$10^{17}$ cm$^{-3}$ has been estimated, whereas in the KI sample the 7.6 eV absorption band was not detected, indicating that oxygen vacancies are in concentration lower than ~$10^{16}$ cm$^{-3}$. Subsequently, the same samples were subjected to a series of isothermal treatments at T=630 K of variable durations. Finally, we note that no care to protect the samples against room light illumination has been taken. In fact, a preliminary investigation we have performed has pointed out that room light has no effects on the concentration of point defects induced in the materials considered.

Optical absorption and luminescence measurements were carried out at room temperature with an ACTON vacuum-UV and a Jasco FP6500 spectrometers, respectively. EPR measurements were carried out at room temperature with a Bruker EMX spectrometer working at frequency $\nu \approx 9.8$ GHz (X-band) and with magnetic-field modulation frequency of 100 kHz acquiring in the first-harmonic unsaturated mode (FH-EPR) or in the high-power second-harmonic mode (SH-EPR). Concentration of defects was determined, with an accuracy of 10%, comparing the double integral of the FH-EPR spectrum with that of a reference sample. For the reference sample the defects concentration was evaluated, with absolute accuracy of 20%, using the instantaneous diffusion method in spin-echo decay measurements carried out in a pulsed spectrometer [15]. The intensity of the second-harmonic EPR signal was estimated by simple integration of the spectra.

After irradiation of the materials considered, $E'_\delta$ and $E'_\gamma$ centers are induced, but virtually no EPR signal due to $E'_\alpha$ centers is detected [7-9]. As discussed extensively in our previous works [7-9], upon thermal treatment at T>500 K the EPR signal of $E'_\gamma$ and $E'_\delta$ centers increases as a consequence of a hole transfer process form the $[AlO_4]^0$ centers to the sites precursors of the E' centers. After prolonged thermal treatments, a third contribution to the EPR spectrum becomes evident, suggesting that $E'_\alpha$ centers are also induced in the same materials by hole transfer [16]. The occurrence of this generation process indicates that the $E'_\alpha$ centers are positively charged. The identification of the three varieties of E' centers was determined by the fit procedure described in Fig. 1(a) for a KUVI sample. The EPR line shapes of $E'_\delta$ and $E'_\alpha$ centers were obtained by the BRUKER SimFonia software, whereas that of the $E'_\gamma$ was determined experimentally in samples in which the EPR signals of $E'_\alpha$ and $E'_\delta$ centers were absent [17]. In Fig. 1(a) the curve obtained as a weighted sum of the three E' centers EPR line shapes is superimposed to the experimental spectrum, for comparison. From the analysis reported in Fig. 1(a), and fixing $g_{||}$=2.0018 for $E'_\gamma$ [1], we obtained the principal g values of the $E'_\alpha$ center reported in Table 1. The comparison of the g values estimated in the present work with those reported previously [1], also shown in Table 1, points out a good agreement for $g_1$ and $g_3$, whereas a less good one is found for $g_2$. In principle, this discrepancy could suggest that the orthorhombic line shape detected in the present work is due to a variant of the $E'_\alpha$ center distinguishable with respect to that reported by Griscom [1, 3, 6]. However, in previous experimental investigations a dependence of the g matrix on the method of defect generation for the $E'_\delta$ [18] and on the γ-ray dose [17] and thermal history of the sample for the $E'_\gamma$ [3,17] centers, have been reported. For this reason we assume that the discrepancy observed in the $g_2$ value of the $E'_\alpha$ center estimated in the present work with respect to that reported by Griscom [1] are property common to the E' centers, presumably reflecting an intrinsic structural relaxation freedom of a-$SiO_2$.

In order to investigate the $^{29}$Si hyperfine structure of the $E'_\alpha$ center, we performed SH-EPR measurements over an expanded field scan. We have found that in the samples in which the $E'_\gamma$ main resonance is detected and that of the $E'_\alpha$ is absent, a pair of lines split by 42 mT is observed, whose spectroscopic features coincide with those attributed to the hyperfine structure of the $E'_\gamma$ center. At variance, as shown in Fig. 1(b), in the samples in which both $E'_\gamma$ and $E'_\alpha$ main EPR lines are detected, the hyperfine spectra show a composite nature. In order to account for these features the hyperfine spectra were fitted comparing the experimental spectrum with a weighted sum of two pairs of lines split by 42 mT and 49 mT, also shown in Fig. 1(b). The 42 mT doublet considered in the fit was obtained experimentally in a sample of EQ906 [13] irradiated at a dose of 5 kGy (in which $E'_\gamma$ but no $E'_\alpha$ centers are induced), whereas the pair split by 49 mT was obtained from the fitting residual of the experimental spectrum of Fig. 1(b) with the 42 mT doublet alone. Once the line shapes of the 42 mT and 49 mT doublets were determined, they were used to fit the experimental hyperfine spectra obtained in all the other materials considered.

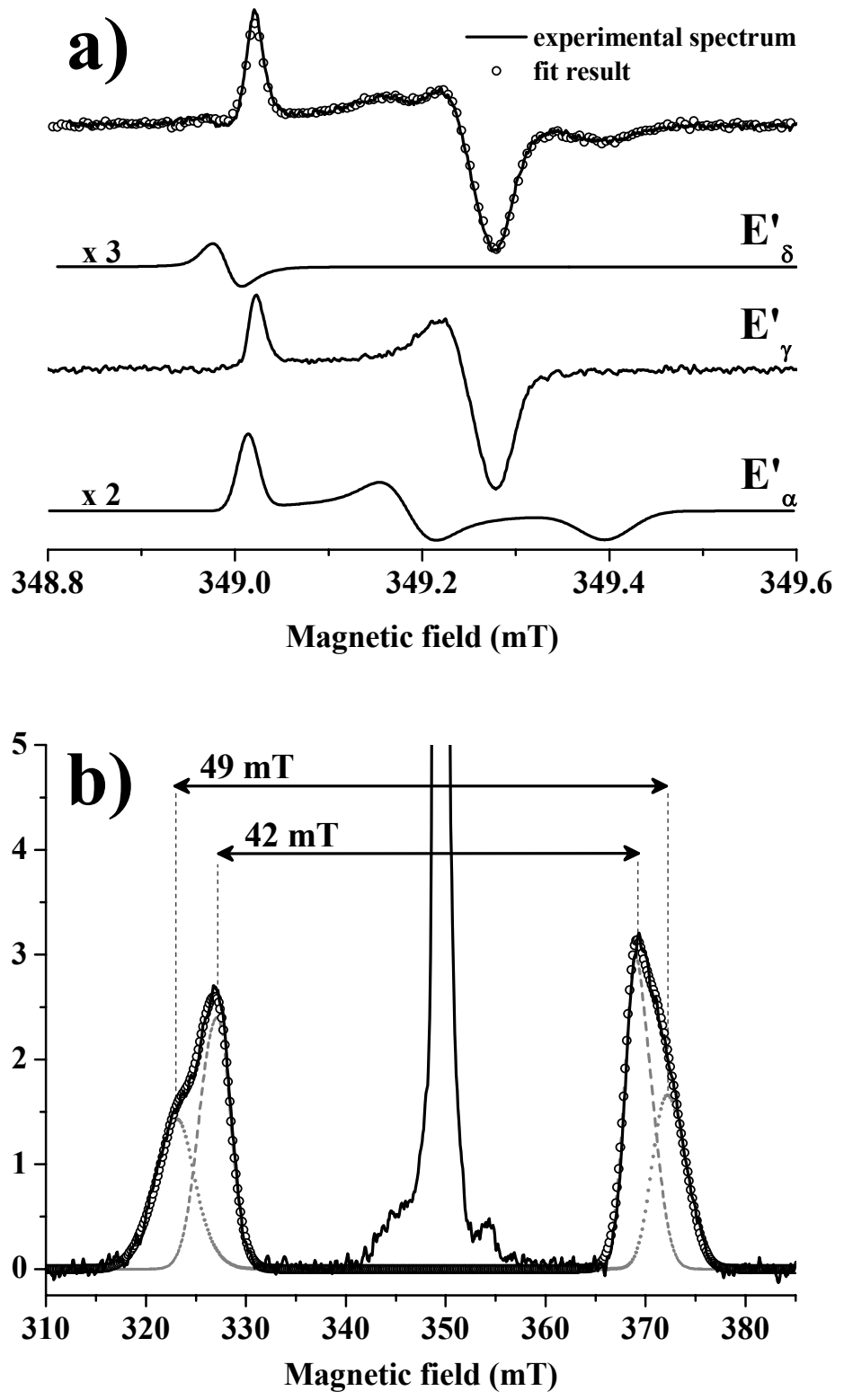

**FIG. 1.** KUVI sample irradiated at ~$10^2$ kGy and isothermally treated at T=630 K for ~$7 \times 10^4$ seconds. (a) Top, FH-EPR spectrum (continuous line) compared to the line obtained as a weighted sum (circles) of the reference lines for $E'_\delta$ , $E'_\gamma$ and $E'_\alpha$ centers, shown in the bottom. (b) SH-EPR spectrum (continuous line) over an extended range compared to the line obtained as a weighted sum (circles) of the reference lines for the ~42 mT and ~49 mT doublets (broken lines).

In Fig. 2(a) we report the concentration of $E'_\gamma$ and $E'_\alpha$ centers in the KUVI sample as a function of the duration of the isothermal treatment [19]. In the same figure the SH-EPR intensities of the 49 mT and of the 42 mT doublets are also reported and point out that the former correlates with the concentration of $E'_\alpha$ center, whereas, as expected, the latter correlates with the concentration of $E'_\gamma$ center. The SH-EPR intensity of the 49 mT doublet was also estimated in QC, Pursil 453 and KI samples thermally treated for various times at 630 K and it is reported in Fig. 2(b) as a function of the $E'_\alpha$ center concentration. As shown, the SH-EPR intensity of the 49 mT doublet and the concentration of the $E'_\alpha$ center change in a strictly correlated way, supporting the attribution of the 49 mT pair to the hyperfine structure of the $E'_\alpha$ center, originating from the hyperfine interaction of the unpaired electron with a $^{29}$Si nucleus. This assignation is also supported by the estimated FH-EPR intensity ratio between the 49 mT doublet and the main resonance line of about ~ 5 %, in agreement with the ~4.7 % natural abundance of $^{29}$Si nuclei. We have estimated that the low- and the high-field components of the 49 mT doublet have a full width at half maximum (FWHM) of ~4 mT and ~4.5 mT, respectively. These values are about 25 % larger than those of the 42 mT doublet components, suggesting a wider distribution of the isotropic hyperfine constant. Furthermore, we have found that the shift of the centroid of the 42 mT with respect to the main position of the $E'_\gamma$ central line is $\Delta g=0.0071\pm0.0005$, and that of the 49 mT with respect to the $E'_\alpha$ is $\Delta g=0.0096\pm0.0005$. Second order hyperfine interaction corrections [20] predict a shift of $\Delta g\cong0.0072$ for the 42 mT and of $\Delta g\cong0.0099$ for the 49 mT doublets, in quite good agreements with the experimental values. The different shift observed for the two hyperfine doublets being due to the dependence of the second order corrections on the square of the hyperfine splitting.

**Table 1. EPR parameters of the $E'_\alpha$ center**

| Point defect | $g_1$ | $g_2$ | $g_3$ | Hyperfine splitting |
|---|---|---|---|---|
| $E'_\alpha$ (Ref. 1) | 2.0018 | 2.0013 | 1.9998 | 42 mT or 13 mT |
| $E'_\alpha$ (this work) | 2.0018 ±0.0001 | 2.0009 ±0.0001 | 1.9997 ±0.0001 | 49 mT |

Although similar annealing features were found in all the materials considered, the maximum concentration of E' centers induced by thermal treatments was found to depend on the material. In fact, the concentration of E' centers induced in the KI sample was found to be more than one order of magnitude lower than that induced in KUVI, QC and Pursil 453, reflecting the lower oxygen deficiency of the KI with respect to the other materials considered. This result suggests that the site precursor of the $E'_\alpha$ defect is oxygen-deficient. Furthermore, we have found that the concentration of twofold coordinated Si defect in the KUVI sample before thermal treatment is about one order of magnitude lower than the concentration of the thermally induced $E'_\alpha$ centers, indicating that the twofold coordinated Si does not act as precursor defect for the $E'_\alpha$ center. Consequently, we suggest that the $E'_\alpha$ center arises from an oxygen vacancy by trapping a hole in a process similar to that proposed for the $E'_\gamma$ [8,9]. Since we exclude that an isothermal treatment at T=630 K could be able to move an oxygen atom out from its regular site of a-$SiO_2$, as suggested in the model proposed by Griscom [3] for the $E'_\alpha$, we suppose that oxygen vacancies present in the material before thermal treatment at T=630 K are involved.

The strict similarity of the $E'_\gamma$ and $E'_\alpha$ centers hyperfine structures suggests that similar Si-$sp^3$ hybrid orbitals are involved in the two defects. On the other hand, the orthorhombic components of the $E'_\alpha$ center g tensor could indicate that a weak interaction of the unpaired electron with the atoms disposed close to the defect occurs. In order to take into account these two properties, we suggest that the microscopic structure of the E'$_\alpha$ center could consist in a hole trapped in an oxygen vacancy with the unpaired electron $sp^3$ orbital pointing away from the vacancy in a back-projected configuration and interacting with an extra oxygen atom of a-$SiO_2$ matrix. The existence of this structure in a-$SiO_2$ has been previously proposed on the basis of experimental [4] and theoretical [5] studies. In particular, in the latter study on the basis of embedded cluster calculations an hyperfine constant of ~48.9 mT [5] was calculated, in excellent agreement with our experimental estimation.

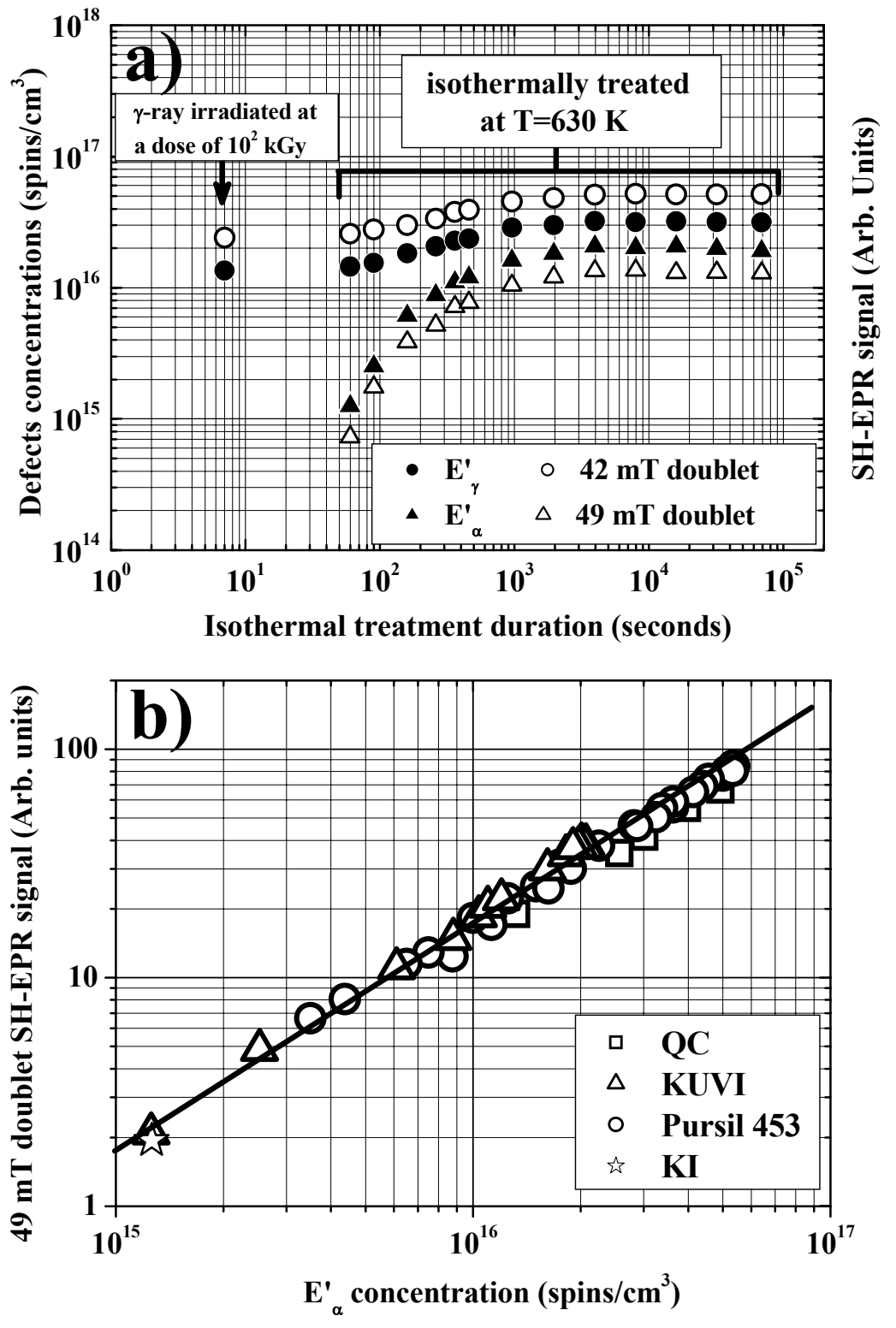

**FIG. 2.** (a) Effects of isothermal treatment at T=630 K on the sample of KUVI γ-ray irradiated at a dose of ~$10^2$ kGy. Filled symbols refer to the left scale, whereas open symbols refer to the right scale. (b) SH-EPR intensity of the 49 mT doublet as a function of the concentration of E'$_\alpha$ centers in four different materials. The straight line, with slope 1, is superimposed to the data, for comparison.

In conclusion, in the present Letter we have shown that the $^{29}$Si hyperfine structure of the E'$_\alpha$ center actually consists in a pair of lines split by 49 mT. Our experimental results support a model for the E'$_\alpha$ center consisting in a hole trapped in an oxygen vacancy with the unpaired electron $sp^3$ orbital pointing away from the vacancy in a back-projected configuration and interacting with an extra oxygen atom of a-$SiO_2$ matrix [4,5].

We thank R. Boscaino's research group, V. Radzig, A. Shluger, A. Kimmel and P. Sushko for useful discussions and suggestions, E. Calderaro and A. Parlato for taking care of the γ irradiation in the irradiator IGS-3 at the Nuclear Department of Engineering, University of Palermo.

* Email address: buscarin@fisica.unipa.it